\documentclass{ws-procs9x6}

\begin{document}

\title{Relativistic Heavy Ion Physics in the New Millennium}

\author{J.L. Nagle}

\address{University of Colorado, Boulder \\
390 UCB \\ 
Boulder, CO 80309-0390, USA\\ 
E-mail: Jamie.Nagle@Colorado.Edu}

\maketitle

\abstracts{
The field of relativistic heavy ion physics has seen significant advancement 
in the new millennium toward a greater understanding of 
QCD at high temperatures with the commissioning and operation of the Relativistic Heavy Ion Collider (RHIC)
starting in 2000.  Here we review progress in the field as presented
in a set of lectures at the Lake Louise Winter Institute on Fundamental Interactions in February 2004.
}

\section{Introduction}
These proceedings are an overview from a set of lectures on the subject of relativistic heavy ion 
physics as given at the Lake Louise Winter Institute on Fundamental Interactions in February 2004.  
These proceedings are meant to highlight progress and open questions in the field, and are no where
near a comprehensive review of the field as a whole.  
We begin with a discussion of the goals of the field and a definition of the quark-gluon plasma.  We 
discuss the role of the quark-gluon plasma in the early universe and in relativistic heavy ion
reactions.  We discuss in detail recent heavy ion experimental results including initial state
physics, bulk particle production, collective motion, and partonic probes of the medium.  We
then summarize with an outlook toward the future.

A number of interesting new results in this field of physics were presented at the 
International Conference on Quark Matter in January 2004.  
All of the conference talks are posted on the web at
\begin{itemize}
\item{http://www.lbl.gov/nsd/qm2004/program.html}
\end{itemize}  
This generated multiple articles in the popular press including three 
in the New York Times all in the same week.\cite{glanz1}
\begin{itemize}
\item {``Newly Found State of Matter Could Yield Insights Into Basic Laws of Nature''}
\item {``Tests Suggest Scientists Have Found Big Bang Goo''}
\item {``Like Particles, 2 Houses of Physics Collide''}
\end{itemize}
The New York Times articles, though certainly simplifying the physics, 
do reflect the excitement of the 
physicists involved, but also the complexity of the system being discovered and the incomplete 
picture we currently have.

\section{Why Are We Here?}

The purpose of the field of relativistic heavy ion physics is to observe and understand 
the nature of Quantum Chromodynamics (QCD) under extreme and novel conditions.  QCD as 
our theory of the strong interaction is full of both fascinating complexity and sometimes 
also amazing simplicity?  It is successful at predicting many experimental observables with 
great precision, and yet unable to be used to directly calculate the basic characteristics 
of hadrons.  Our inability to study in isolation the nature of the fundamental particles 
of the theory (quarks and gluons) presents many challenges to furthering our knowledge.  
What about QCD matter
under extreme conditions of high temperature or high density?  Can we understand 
characteristics of the matter that dominated the very earliest stages of the universe?  
Can we observe characteristics of hot and dense nuclear matter in the laboratory with 
relativistic heavy ion collisions?  Can these observations give us insight about the 
transition from partons bound in hadrons to a deconfined system of quarks and gluons?
These are the questions for which we seek answers.

\section{Quantum Chromodynamics}

\begin{figure}[ht]
\centerline{\epsfxsize=4.1in\epsfbox{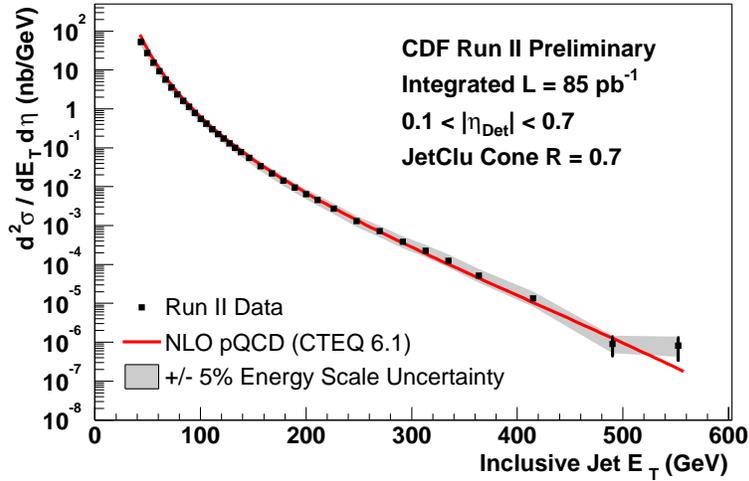}}
\caption{CDF Run II Preliminary Inclusive Jet Yield versus transverse energy compared
with an NLO pQCD calculation using CTEQ6.1 Parton Distribution Functions.\label{fig_cdfjets}}
\end{figure}

Most of us believe that QCD is the correct quantum field theory for strong interactions.  
We believe this because of the successful calculation and experimental verification of, among 
other things, high energy jet yields and the evolution of the 
partonic structure of the proton as calculated via DGLAP evolution.\cite{dokshitzer}
Figure~\ref{fig_cdfjets} shows transverse energy jet spectra from CDF in Run II at the 
Tevatron compared quite favorably with an NLO pQCD calculation.\cite{cdf}

\begin{figure}[ht]
\centerline{\epsfxsize=3.1in\epsfbox{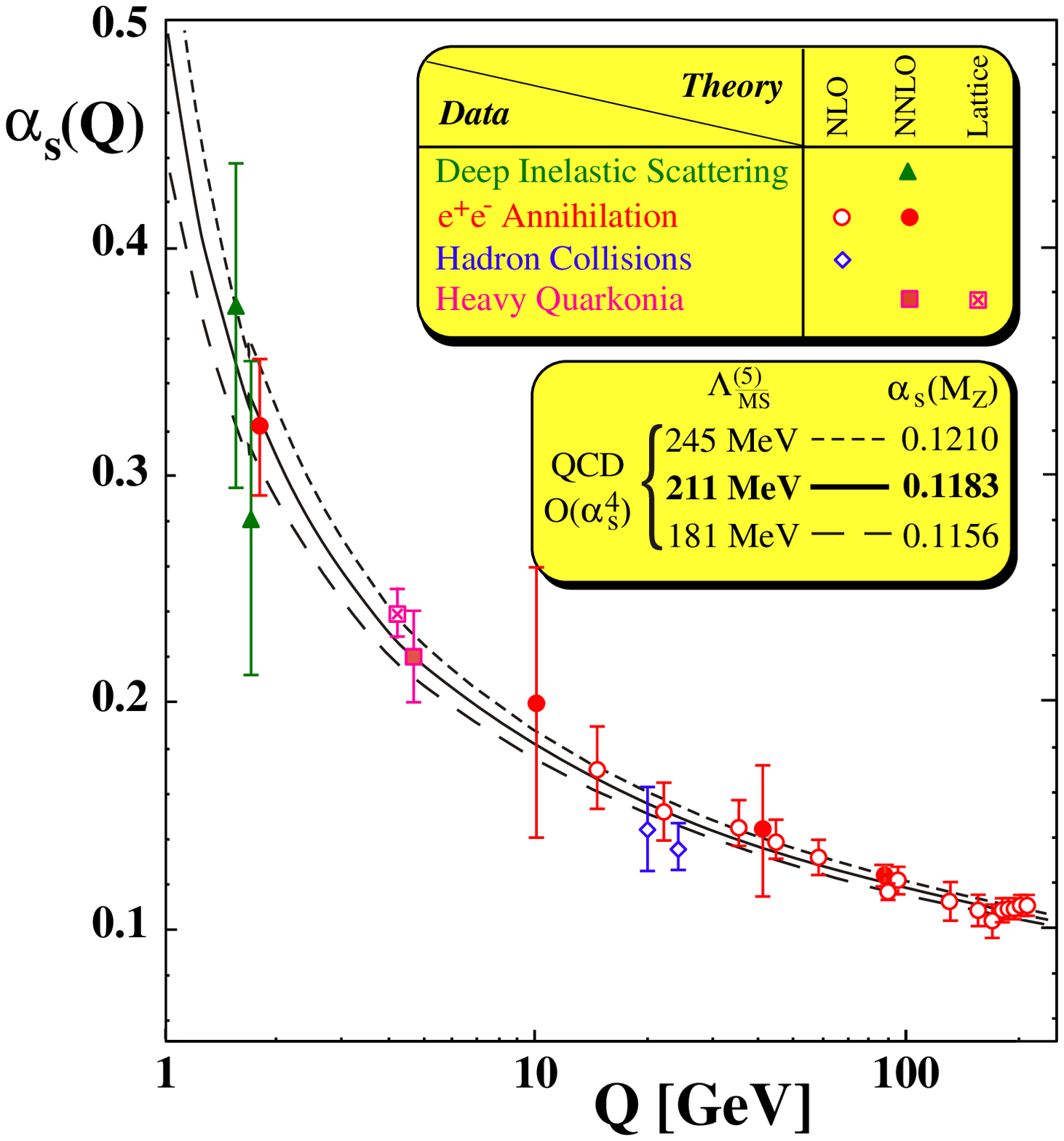}}
\caption{The strong coupling $\alpha_{s}$ as a function of $Q$ in $GeV$ as measured
in various channels.\label{fig_alphas}}
\end{figure}

\begin{figure}
\centerline{\epsfxsize=3.1in\epsfbox{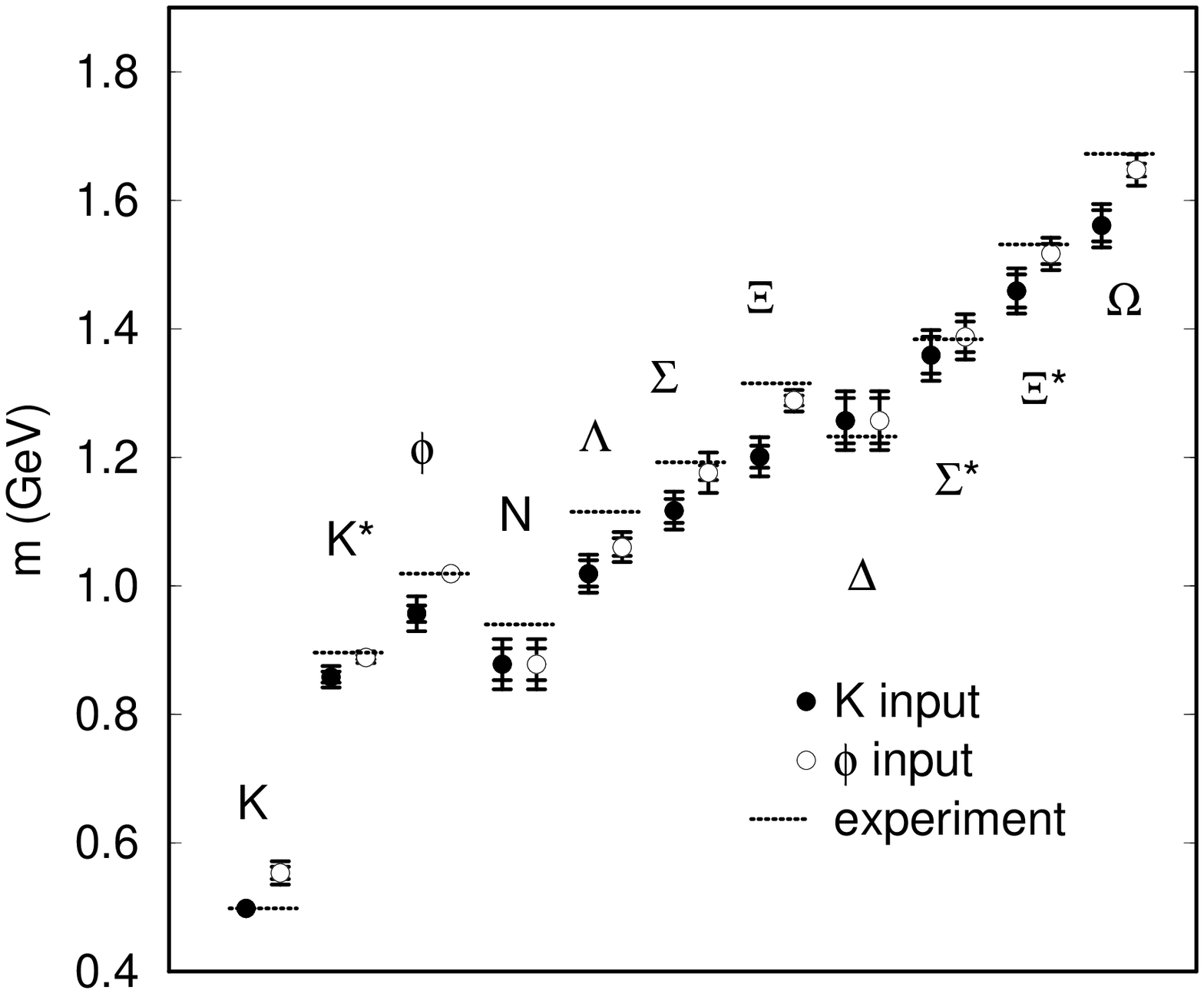}}
\caption{Hadronic masses calculated via lattice QCD compared with the
experimentally determined spectra.\label{fig_latticemass}}
\end{figure}

These examples are next-to-leading order perturbative QCD calculations that are 
applicable at large $Q^{2}$.  
However, one cannot use the same calculation technique to determine the absolute partonic 
structure of hadrons, only their evolution in $Q^{2}$.  
Figure~\ref{fig_alphas} shows the strong coupling $\alpha_{s}$ as a function of $Q$.  At 
large $Q$, $\alpha_{s}$ is small and the perturbative expansion is convergent, while at 
low $Q$ this is no longer the case.
So what about the non-perturbative 
world around us?  Using numerical techniques of lattice QCD one can calculate the various 
hadron masses and obtains agreement at around the 10\% level
(with the possible exception of the Goldstone boson pions), as
shown in Figure~\ref{fig_latticemass}.\cite{latticemass}

If we want to ask questions about the nature of partonic matter at high energy densities, 
we are often forced to rely on lattice QCD calculations.  However, it should be noted that 
at truly high temperatures (far above what we can achieve with RHIC or at the future LHC), 
the average $Q^{2}$ for the individual parton-parton scattering can be quite large.  
This ``perturbative'' plasma would have only ``weakly'' interacting quarks and gluons 
as their interactions would be dominated by the region where $\alpha_{s}$ is small.  
However, at temperatures and energy densities more realistic to our experiments 
($T \approx 100-500~MeV$), perturbative calculations are sure to break down.

\section{Quark-Gluon Plasma}

\begin{figure}
\centerline{\epsfxsize=4.1in\epsfbox{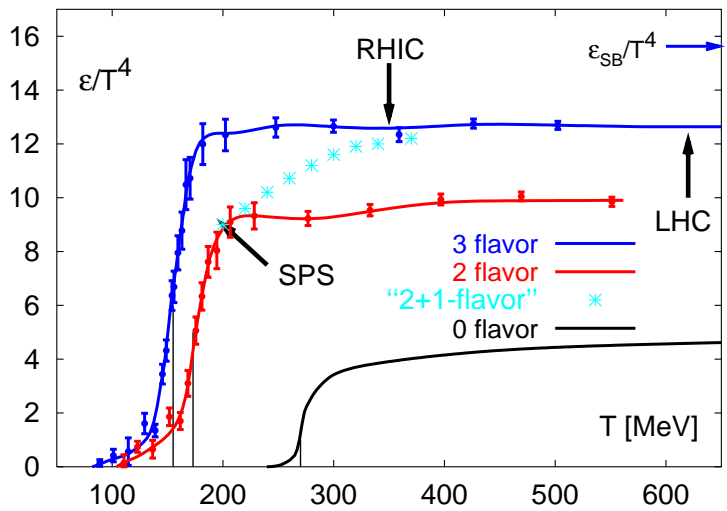}}
\caption{Lattice QCD results for the energy density / $T^4$ as a function of
the temperature ($MeV$).  Note the arrow on the right side indicating the
level for the Stefan-Boltzmann limiting case.\label{fig_lattice}}
\end{figure}

Lattice QCD predicts a phase transition to a quark-gluon plasma at a temperature of
approximately $T \approx 170~MeV \approx 10^{12}$ Kelvin, 
as shown in Figure~\ref{fig_lattice}.\cite{karsch}  
This transition temperature corresponds to an energy density $\epsilon \approx 1~GeV/fm^{3}$, an order
of magnitude larger than normal nuclear matter density.
Calculations indicate a significant change in behavior of the system over a small change in temperature 
including restoration of approximate chiral symmetry.  
Energy densities above the transition value correspond to many hadrons per cubic fermi.
Hadrons cannot exist as in vacuum at this density.  
There is no way for a parton to know which hadron wavefunction it belongs to at this
density.

The exact order of this phase transition is not known.  In a pure gauge theory (only gluons) 
the transition appears to be first order.  However, inclusion of two light quarks 
(up and down) or three light quarks (adding the strange) can change the transition 
between 1st order, 2nd order and a smooth crossover.  These calculations are at zero 
net baryon density and the nature of the transition and the medium itself may significantly 
change as one changes the net baryon density of the medium.  
 
In the case of massless non-interacting particles, one has a simple relation between
the energy density $\epsilon$ and the temperature $T$ depending on 
the number of degrees of freedom in the system $g$.
\begin{equation}
\epsilon = g {{\pi^{2}} \over {30}}T^{4}
\end{equation}
In the case of temperatures of order $T \approx 100-500~MeV$, in a hadron gas there are three
degrees of freedom for the three Goldstone boson pions.  However, in a quark-gluon plasma there
are over 30 degrees of freedom from the light quarks and gluons.

It is interesting to note that the lattice results indicate a system that is still
significantly different from the Stefan-Boltzmann limit.  
However, many people state that the lattice is ``not so far from the non-interacting gas limit'', 
and thus expect a ``weakly'' interacting gas of quarks and gluons.  
Walter Greiner states that 
``in order to allow for simple calculations the QGP is usually described as a free gas consisting of quarks and gluons.  This is theoretically not well founded at $T \approx T_{c}$.''\cite{greiner}
The quasi-particles in the plasma may be phonons or plasmons rather than quarks and gluons.  
The plasmons would arise because near the transition temperature the effective 
coupling $\alpha_{s}$ could be large and a dynamical mass $m_{g} \approx T_{c}$ could be 
generated by gluons.

\begin{figure}
\centerline{\epsfxsize=3.1in\epsfbox{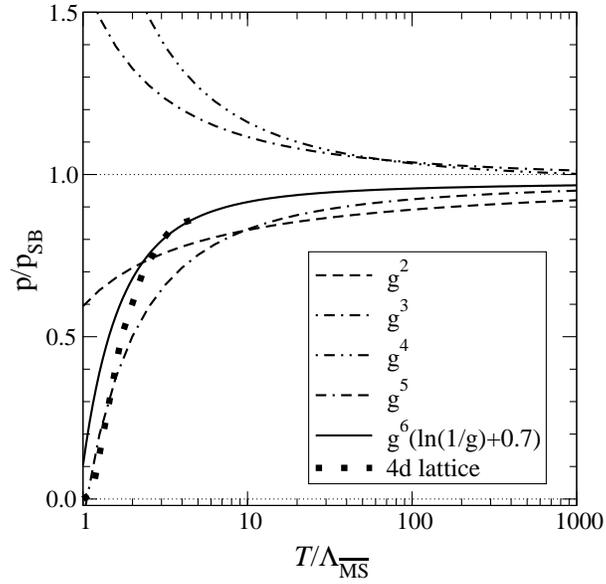}}
\caption{Perturbative QCD results for the pressure as a function of temperature 
at various orders normalized to the Stefan-Boltzmann value $p_{SB}$.\label{fig_hotqcd}}
\end{figure}

How high a temperature is needed not just to form a quark-gluon plasma, but to approach this
``weakly'' interacting plasma?  
A calculation of the pressure of hot matter within perturbative QCD is shown in 
Figure~\ref{fig_hotqcd}.\cite{hotqcd}
The pressure result oscillates significantly as one considers contributions of different orders.
These oscillations are an indication that the expansion is not yielding reliable results.  However
at temperatures approaching 1000 times $T_{C}$, they appear to be converging toward the
Stefan-Boltzmann limit (asymptotically free partons).  It is interesting that in considering
the highest order term, the results are still non-convergent though one seems to approach the
lattice calculated pressure.  Due to the high parton occupation, the perturbative expansion is
never completely convergent as one gets in single parton-parton scattering.

\begin{figure}[ht]
\centerline{\epsfxsize=4.1in\epsfbox{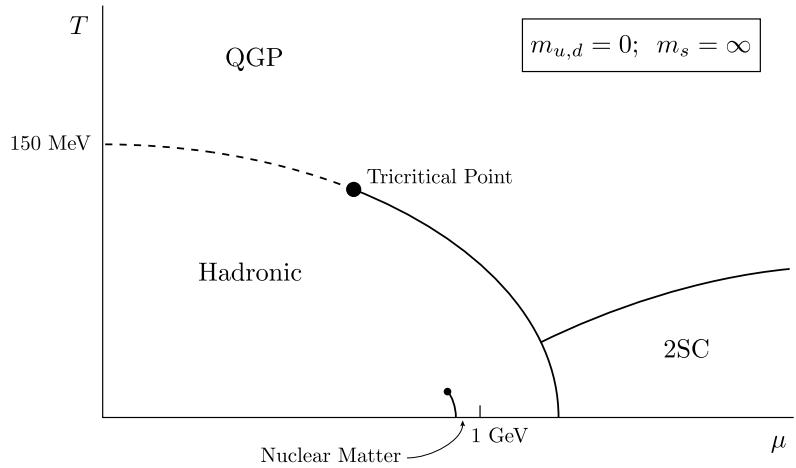}}
\caption{Theoretical phase diagram of nuclear matter, not yet confirmed by experiment.\label{fig_phase}}
\end{figure}

A schematic version of a phase diagram of nuclear matter is shown 
in Figure~\ref{fig_phase}.\cite{krishna}  This figure shows a transition to
a color superconducting phase of matter at large density and low temperature.  In this phase 
one has partons forming Cooper pairs.  At high temperature and low density, the latest
lattice results with a realistic strange quark mass indicate, though not definitively, that
the transition is a smooth crossover.  Thus, there may be a tricritical point in the
phase diagram connecting a first order transition at high density to this crossover.

Where can we study nuclear matter under such extreme conditions where we might hope to have 
advancement in QCD theory and experimental access?  The early universe shortly after 
the big bang is a system where one almost certainly started out with a ``perturbative'' 
plasma which cooled into something we might try to describe on the lattice, 
and then transitioned into hadrons.  Neutron stars are a system with very high 
net baryon density (of order ten times that of normal nuclear matter), but very low temperature and
thus might be relevant for studying the color superconducting phase.  
And then there is the relativistic heavy ion program where we can control the system in
the laboratory.

\section{Transitions of the Early Universe}

In our current understanding of the history of the universe, shortly after the big bang, the 
universe went through an enormous expansion during an inflationary period.  Post-inflation, 
the radiation dominated universe gave way to a universe filled with quark-gluon plasma.  Six 
microseconds after the big bang, the universe cooled below a temperature of approximately 
$200~MeV$ and the quark-gluon plasma had all the quarks and gluons confined into hadrons.  
One second later, at a temperature of order $2~MeV$, the hadrons (protons and neutrons) formed 
light nuclei in a process referred to as big bang nucleosynthesis (BBN).  
Then 300,000 years later, electrons and nuclei (mostly protons) formed atoms, at which 
point the universe became relatively transparent for photons.  
Photons at this point largely decoupled from the universe and are the cosmic microwave background 
we observe today.  

We are particularly interested in the transition from a bath of quarks and gluons 
(quark-gluon plasma) to a system of hadrons (for example protons and neutrons) where the 
quarks and gluons (partons) are bound.  

In a seminal paper, Ed Witten wrote in 1984 that 
``a first-order QCD phase transition that occurred in the early universe would 
lead to a surprisingly rich cosmological scenario,''  and ``although observable 
consequences would not necessarily survive, it is at least conceivable that the
phase transition would concentrate most of the quark excess in dense, invisible 
quark nuggets.''\cite{witten}
These quark nuggets could be in the form of strange quark matter (SQM), composed of roughly 
equal numbers of up, down and strange quarks.  Bubbles of supercooled quark-gluon plasma 
could have formed strange quark matter nuggets with baryon number ($A >> 1000$).  

Strange quark matter could even be more stable than $Fe^{55}$ and thus be the true ground state of 
nuclear matter.  Iron would then decay into SQM, but with a lifetime longer than the age 
of the universe since it would be a 55th order weak decay.  If SQM were not just metastable, 
but completely stable, it could be a source of baryonic dark matter.  Despite the heavier 
neutral current mass of the strange quark (relative to up and down), the new flavor quantum 
number may allow for a lower total energy if the Fermi energy levels for up and down quarks 
are occupied.  Twenty years after Witten's paper many searches for 
strange quark matter have been made in terrestrial matter, nuclear reactions, and in 
relativistic heavy ion collisions.  All have yielded null results to date.\cite{sqm,e864}

Also, if the plasma-to-hadrons transition were a strong first order phase transition, bubble formation 
in the mixed phase could have resulted in a very
inhomogeneous early universe.  Big bang nucleosynthesis calculations assume a 
homogeneous universe.  This line of investigation was quite active when the dark matter issue 
raised questions about the implied baryon content of the universe as derived from big bang 
nucleosynthesis.  An important question is whether inhomogeneities from bubbles in the
mixed phase would survive diffusion as the universe cooled down to $2~MeV$ when BBN occurred?
Are the bubbles too small and close together such that diffusion erases the inhomogeneities
before nucleosynthesis?

In 2001, the Boomerang Experiment reported that
``The value deduced from the second harmonic in the acoustic oscillations for 
$\Omega_B$ = $0.042 \pm 0.008$ (cosmic baryon mass density) is in very good 
agreement with the value one gets by applying the theoretical details of primordial 
big bang nucleosynthesis to the observations of cosmic abundances of deuterium.''\cite{boomerang}
Does this observation rule out a first order phase transition in QCD?
This confirmation of BBN does not rule out a first order phase transition in QCD because of 
the diffusion issue, but places some model dependent limit on the level of supercooling and
bubble formation in the transition.  

The universe at 300,000 years old is extremely homogeneous, isotropic to one part in 100,000.  
The WMAP experimental results now provide an amazing advance in our precision knowledge of 
the early universe, though still little constraint on the quark-gluon plasma phase.  
Unfortunately nothing decoupled at the QCD transition for us to still measure today, 
like the photons 300,000 years later when atoms were formed.  Thus, the study of extreme QCD in this
regime must turn to accelerators rather than the sky.

\section{Relativistic Heavy Ion Collider}

The Relativistic Heavy Ion Collider (RHIC) facility is located at Brookhaven National 
Laboratory in Upton, New York, USA.  The facility was first commissioned and brought 
on-line in 2000.  Heavy ion reactions have been studied over the last three decades
at ever increasing energy.  
At the lowest energies, the reactions exchange protons and neutrons and one can 
study issues of nuclear structure.  At intermediate energies, the nuclei are largely 
broken apart and one has a cascading of nucleons and excitation of hadronic resonances.  
At the Brookhaven AGS and CERN SPS fixed target programs, one hoped to create the quark-gluon 
plasma state, albeit with higher net baryon density and lower temperature than at RHIC.  Most 
of this presentation focuses on the recent RHIC results, but it should always be kept in 
mind how these results connect and build upon lower energy results.  Certainly as the Large 
Hadron Collider (LHC) heavy ion program begins in 2007, the connections between RHIC and 
LHC will be a critical tool for understanding QCD matter.

The first gold-gold collisions at RHIC were observed by all experiments in June 2000 
at $\sqrt{s}_{NN} = 130 ~GeV$.  There was immediately some ``interesting'' press coverage 
about possible disaster scenarios if RHIC created a black hole.  As a simple exercise, 
the Schwarzschild radius for all the energy deposited in a RHIC collision would be $10^{-49}$ 
meters, which is much smaller than the Planck length!  In comparison, the compression in the 
reaction and subsequent expansion is never expected to have a system smaller than $10^{-16}$ 
meters.  If we ignore the conflict of quantum mechanics and general relativity at the Planck 
scale, we find the black hole, even if somehow there is over three orders of magnitude more 
compression than expected, would evaporate by Hawking radiation in $10^{-83}$ seconds.  

RHIC has been quite successful in its commissioning of the heavy ion program and in 2004 
(the fourth running year of the collider), the design energy $\sqrt{s_{NN}} = 200 ~GeV$
and design luminosity for gold-gold 
collisions was achieved, and more importantly, successfully run over many months of data 
collection.  As we stated at the conference ``RHIC is kicking butt!''

\section{Experiments}

There are four dedicated experiments in the RHIC program for heavy ion studies.  The 
two large scale experiments are STAR and PHENIX.  The STAR experiment focuses on 
hadronic observables over a very large acceptance.  Their detector consists of a 
solenoid magnetic field and a large coverage time-projection chamber (TPC).  The 
TPC is the primary tracking device and is augmented by an inner silicon detector, 
ring-imaging Cerenkov detector, electromagnetic calorimeter and time-of-flight system.  The 
PHENIX experiment focuses on electromagnetic probes of the medium.  The detector is 
comprised of four separate spectrometers.  Two at 90 degrees with tracking detectors, 
ring imaging Cerenkov counter, and electromagnetic calorimetry all used to identify 
photons, electrons and hadrons.  Two spectrometers at forward angles are designed 
to track muon primarily for reconstruction of quarkonia states.

The two smaller experiments are BRAHMS and PHOBOS.  BRAHMS has two rotating spectrometer 
arms with excellent hadron particle identification detectors.  Their spectrometers cover 
the broadest range of rapidity for identified hadrons.  PHOBOS characterizes the reactions 
over nearly $4\pi$ with silicon detectors.  Additionally they have a central spectrometer 
to measure charged hadrons in many silicon layers, augmented with a time-of-flight 
scintillator system.

\section{Initial State Physics}

\begin{figure}[ht]
\centerline{\epsfxsize=3.1in\epsfbox{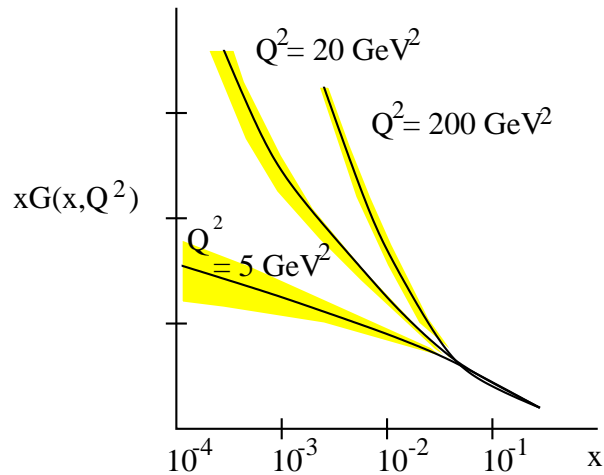}}
\caption{Gluon Structure $xG(x,Q^{2})$ as a function of x for three values
of $Q^{2}$.\label{fig_q2evolution}}
\end{figure}

In this section we review the relevant physics of the initial state, before the two 
nuclei collide and immediately afterwords.  
Deep inelastic scattering (DIS) experiments at HERA have 
revealed an amazing complexity of the partonic structure of the proton.  At short wavelength 
(large values of $Q^{2}$) there is an enormous growth of low x gluons in the proton's 
structure.  Note that x is the fraction of the proton's momentum carried by the parton.  
In fact, at high $Q^{2}$ the proton is almost completely dominated by this sea 
of gluons as shown in Figure~\ref{fig_q2evolution}.\cite{larry}

When we collide heavy gold ions at RHIC, we essentially ignore the nuclear shell structure 
for nucleons since the kinetic energy of the reaction easily overwhelms the nuclear binding energy.  
Since the energy is quite large, one can think of the collision as being between two walls of gluons.  
In that sense, some people have referred to RHIC as a gluon collider.  The nuclei are 
merely containers for transporting these gluons to the collision point.

In the collision, of order 10,000 gluons, quarks and anti-quarks (mostly gluons) 
are freed from their virtual excitation in the nuclear wavefunction and made physical 
in the laboratory.  What is the nature of this ensemble of partons?  One thing for certain 
is that it represents a major technology and detector challenge to measure and characterize 
these reactions.  The RHIC experiments and community have successfully met this challenge 
and the detectors have all worked remarkably well in this high particle density environment.

\subsection{Particle Multiplicity}

In collisions of electrons and positrons ($e^{+}e^{-}$), the annihilation can yield 
back to back outgoing quark and anti-quark partners.  Due to the confining nature of QCD, these 
partons fragment into hadrons, collectively referred to as jets.  One can ask the question of how 
this hadronization process results in a certain multiplicity of hadrons and the energy 
distribution among these hadrons.  At reasonably high energy one can calculate 
perturbatively the evolution of radiated gluons by this quark and anti-quark.  If one then assumes 
parton-hadron duality (one scale factor), one can predict the collision energy dependence 
of the hadron multiplicity.  Calculations by Mueller and others reveal an excellent 
agreement with the experimental data from $\sqrt{s}$ of 10 to 200 $GeV$.\cite{mueller}

In proton-proton and proton-antiproton reactions, the hadron multiplicity is noticeably 
lower than in $e^{+}e^{-}$ reactions at the same $\sqrt{s}$.  One can understand this 
qualitatively since much of the available energy in proton-proton reactions is carried 
in longitudinal motion, often referred to as incomplete stopping or the leading particle
effect.  In fact, if one 
scales the $\sqrt{s}$ down to account for the energy in the forward direction, one sees 
reasonable agreement between the $e^{+}e^{-}$ and proton-proton reactions.\cite{ppscale}  
This is a surprising scaling observation since in the $e^{+}e^{-}$ case one can calculate the the 
multiplicity perturbatively, but in proton-proton reactions one cannot since their is 
not a set of simple diagrams and gluon radiations.  

\begin{figure}[ht]
\centerline{\epsfxsize=3.5in\epsfbox{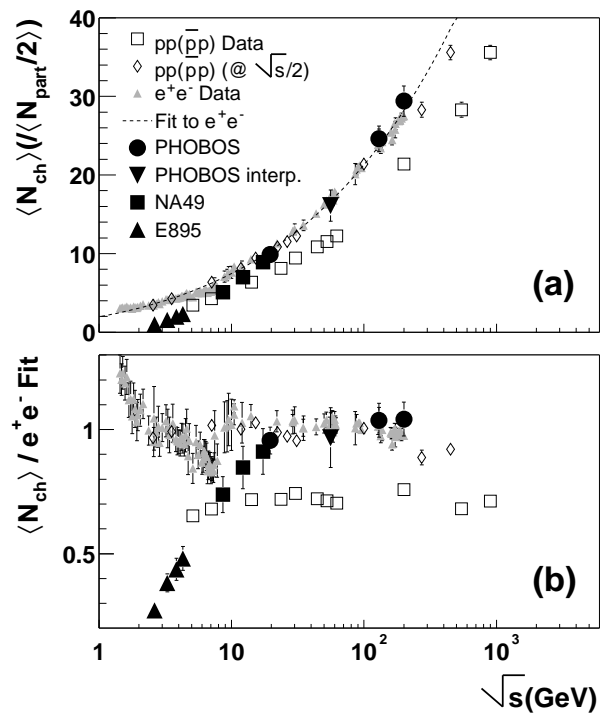}}
\caption{Charged particle multiplicity per participant pair as a function
of $\sqrt{s}$ for proton-(anti)proton and heavy ion gold-gold data.  Also shown
are the $e^{+}e^{-}$ data and a best line fit.\label{fig_ee}}
\end{figure}

The PHOBOS experiment has checked 
this scaling now including heavy ion reactions.\cite{phobos_ee}
As shown in Figure~\ref{fig_ee}, one observes a reasonable similarity 
in the energy dependence.  What does this mean?  It is difficult to reconcile since if 
the only thing that mattered was the total energy deposition per unit volume, perhaps 
this scaling is expected.  However, it is claimed that $e^{+}e^{-}$ does not follow 
a statistical distribution, but rather the perturbative gluon radiation distribution.  
This is an active discussion, but for example the large difference in multiplicity between gluon
and quark jets implies that not all energy is equally available for charged particle production.

These global observations are puzzling.  One might ask if $e^{+}e^{-}$ reactions are then the same 
as gold-gold.  The clear answer is no, but it may be that the mechanism for deciding 
how energy is distributed between particle multiplicity and kinetic energy per particle 
reflects some simple underlying physics.

\subsection{Proton and Nuclear Parton Structure}

One theoretical framework to attempt to understand the multiplicity of particles, 
is by viewing particle production as freeing virtual partons from the incoming 
nuclear wavefunction.  As stated previously, the proton and nucleus are dominated 
by low x gluons when viewed at high $Q^{2}$.  One of the great successes of QCD is 
the agreement of NLO DGLAP evolution over a broad range of $Q^{2}$.\cite{dokshitzer}  
However, the DGLAP fits have many free parameters (18-30) and the form of the parameterization 
is not given by theory.  In fact, for $Q^{2}$ of order $1~GeV^{2}$, the gluon density is 
going negative.  This is physically allowed but may still be a hint of a breakdown in DGLAP.
One problem in interpreting this data is that 
for the HERA experiments the lowest x values probed are always at the lowest $Q^{2}$, 
where the breakdown of NLO DGLAP may not be surprising.  The current HERA running 
may not resolve these issues since machine changes limit the coverage at the lowest 
x values.  Future electron-ion collider (eRHIC) or future running at HERA may be needed.  

One explanation for the breakdown of DGLAP is given in the
context of gluon saturation models.  In deep inelastic scattering (DIS) the physics is 
often described in the target rest frame.  One can imagine the photon as fluctuating 
into a quark and antiquark pair (color dipole).  As one evolves to higher $Q^{2}$ the 
color dipole radiates additional gluons until one approaches the unitary limit of the 
cross section.  It has been shown that this is the equivalent to a very different 
description in the probe rest frame.  Now imagine a probe striking a very 
Lorentz contracted proton or nucleus.  The wavefunction of low x gluons will overlap 
and the self-coupling gluons fuse.  This then saturates the density of gluons in the 
initial state.  Venugopalan, McLerran and collaborators show that in this 
limit, factorization breaks 
down and one can describe the proton or nucleus in terms of classical gluon fields as 
solutions of the Yang-Mills equation.\cite{larry}  
These solutions are often referred to as the color glass condensate (CGC).  
Note that while one could potentially make a physical system 
in this saturation regime, in the nuclear wavefunction the CGC is like a Fock state of the 
wavefunction, and so not the same as a state of matter like the quark-gluon plasma. 

It is now well established experimentally that the nucleon structure functions are modified in nuclei.  
There are various modifications including suppression of partons at high x (EMC effect), 
possible enhancement of partons at intermediate $x \approx 10^{-1}$ (anti-shadowing), and 
large suppression of partons at low $x \approx 10^{-2}$ and below (shadowing and possibly 
saturation).  In the low x regime, one can think of the gluons from all the nucleons 
in the nucleus in a longitudinal slice as overlapping and thus enhancing the gluon 
recombination (more saturation).  
\begin{equation}
xg(x_{eff},Q^{2}) = A^{1/3}xg(x,Q^{2})
\end{equation}
Thus, it may be true that at low $Q^{2} \approx 1 GeV^{2}$ an x value of $10^{-7}$ in the proton 
might be the equivalent of $10^{-2}$ in a nucleus.  This remains to be tested.

These saturation models have been applied to RHIC data and predict the total charged particle multiplicity
distributions with a free parameter for the saturation scale $Q_{sat}$ for gold nuclei.  
It is not directly calculable the relation between the saturation scale in the 
proton measured at HERA and for a gold nucleus measured at RHIC.
Kharzeev and collaborators find quite good agreement with the growth of charged particle 
multiplicity with reaction impact parameter, and also the angular distribution of charged 
particles.\cite{dima}  They assume that the saturation scale drives all the physics and then invoke
parton-hadron duality with a scale factor to compare with experimental results.
This agreement is intriguing, but not yet compelling.  A key question is whether 
this universal behavior of QCD in the saturation limit is the driving physics at 
HERA and RHIC and whether the details to relate the calculations to final
state hadronic observables are well constrained.  Comparison with the
transverse energy production is another important check.\cite{raju}

\begin{figure}[ht]
\centerline{\epsfxsize=4.1in\epsfbox{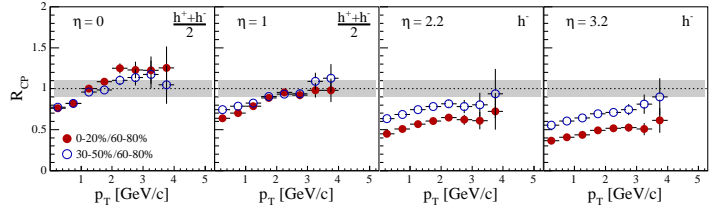}}
\caption{BRAHMS experimental results on charged hadron production in
deuteron-gold reactions for different pseudo-rapidities as a function of
transverse momentum.  The ratio of yields in central to peripheral reactions 
scaled by the nuclear thickness ratio is shown.\label{fig_brahms}}
\end{figure}

One major issue is that at RHIC particle production is dominantly from low $Q^{2}$ processes where
one cannot compare the saturation picture to a firm QCD prediction without saturation.  In addition,
there are expected to be medium effects that can mask the physics.  In Run III at RHIC there was a
study of deuteron-gold reactions where one expects almost no medium effects.  It should be noted that 
this running is nearly equivalent to proton-gold reactions, but with the matched rigidity of the
two beams, the accelerator was able to achieve collisions without having to move the intersection
DX magnets.  If one looks at forward rapidity, one is effectively probing lower x partons in the gold
nucleus.
\begin{equation}
x \approx {M \over \sqrt{s}} \times e^{-y}
\end{equation}
Data from the BRAHMS experiment indicates a significant suppression of hadrons at moderate $p_{T}$
as one measures at more forward rapidity, as shown in Figure~\ref{fig_brahms}.\cite{brahms}
This may be an indication of nuclear shadowing as described via saturation, though more quantitative
comparisons must be made.  Also, this intermediate $p_{T}$ is not free from soft physics
contributions which may complicate the interpretation.  Additionally, the PHENIX experiment has
shown similar data but also at backward rapidity, thus sampling high x partons in the gold nucleus.
This may shed light on anti-shadowing, and perhaps help disentangle the soft and hard
physics contributions.

\section{Collective Motion}

After the initial set of partons are freed from the nuclear wavefunction, either 
via the color glass condensate or otherwise, they have the possibility to interact 
with each other.  Alternatively if they behave as a non-interacting ideal gas, they 
may simply exit and individually hadronize eventually leaving signals in our detectors.  
This would simply appear as a superposition of proton-proton reactions.  

\begin{figure}[ht]
\centerline{\epsfxsize=4.1in\epsfbox{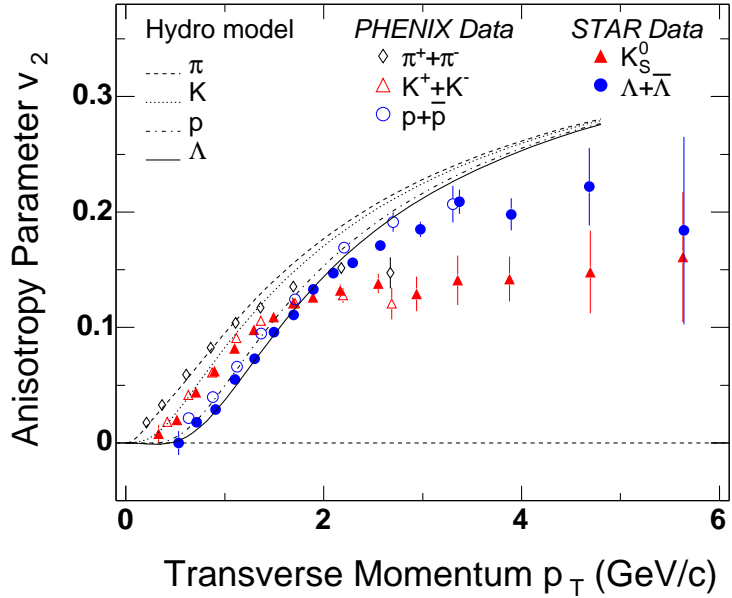}}
\caption{Anisotropy parameter $v_{2}$ as a function of $p_{T}$ for various
hadron species.  Also shown are results from a hydrodynamic model calculation.\label{fig_flow}}
\end{figure}

We have an excellent experimental handle on the coupling of the partons in the system, 
through what is referred to as elliptic flow.  In non-central gold-gold reactions, the 
created system is almond shaped with very different density gradients along the impact 
parameter direction and perpendicular to it.  This large spatial anisotropy  can be 
translated into a momentum space anisotropy if there is a large pressure (microscopically, 
a lot of re-scattering).  Experiments can measure the azimuthal angle distribution of 
particles and make a Fourier decomposition.  The second component $v_{2}$ is referred 
to as the elliptic flow.

As shown in Figure~\ref{fig_flow} there is a large $v_{2}$ that is increasing with transverse momentum 
for all hadronic species measured.\cite{starflow,phenixflow}
Even the multi-strange baryons as measured by the 
STAR experiment appear to have large collective behavior.  PHENIX has shown preliminary data
that give a slight hint that the charm mesons may follow a similar flow pattern, 
though more data needs to be analyzed to confirm such a result.\cite{kelly,batsouli}  

If one assumes that there is enough re-scattering to achieve local equilibration, 
and we assume some initial conditions for the density, one can make use of 
hydrodynamic calculations.  One can use simple equations of motion which are solvable 
using an equation of state, that is derived from lattice QCD.  There is good agreement 
between the experimental data at mid-rapidity for identified hadron transverse 
momentum spectra and also the $v_{2}$ at low transverse momentum $p_{T}<2~GeV$.  It is
notable that the hydrodynamic description fails above this $p_{T}$ which may not be
surprising as there must be a finite scattering limit.  It is also extremely important to note that
the energy density and starting time for the calculations is of order $\epsilon ~\approx
20~GeV/fm ^{3}$ and $t \approx 1 fm/c$.\cite{kolb,ed}

The success of the hydrodynamic calculations with zero additional viscosity (which would 
result from finite mean free paths) has led many to conclude that the system must be 
``strongly'' coupled (i.e. lots of re-scattering).  Molnar and 
collaborators have made a parton cascade calculation 
including only the perturbatively calculable part of the scattering cross 
section.\cite{molnar}  Effectively this is a calculation of a ``perturbative'' ``weakly'' interacting 
plasma of quarks and gluons.  One can think of the partons as having asymptotic freedom 
within the medium.  Most likely this is not applicable to RHIC collisions since $\alpha_{s}$ 
is still large for average parton-parton scatterings, but is an important benchmark.
They find they under-predict the $v_{2}$ for hadrons leading many to conclude that we have
formed a ``strongly'' coupled system, in contrast to the calculations ``weakly'' coupled
perturbative system, which is perhaps not so surprising.  It should be noted that there is a question
of how to map from the partonic $v_{2}$ in the calculation to the hadronic $v_{2}$ that is measured.
This issue needs clarification since a parton coalescence mechanism (as discussed later)
may significantly enhance the resulting hadron $v_{2}$.

\begin{figure}[ht]
\centerline{\epsfxsize=3.1in\epsfbox{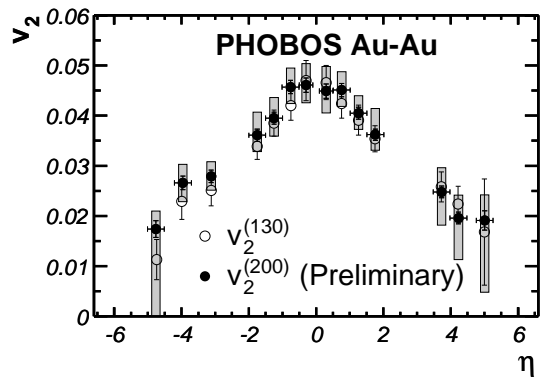}}
\caption{PHOBOS measured $v_{2}$ for charged particles as a function of pseudo-rapidity
in gold-gold reactions at two energies.\label{fig_phobosflow}}
\end{figure}

The hydrodynamic models have trouble describing the longitudinal motion, where it is observed that $v_{2}$ 
falls quite quickly in moving to forward and background rapidity, 
as shown in Figure~\ref{fig_phobosflow}.\cite{phobos_flow}
Many of the models assume boost-invariance, as in the Bjorken expansion scenario, which is not 
supported by the BRAHMS rapidity distribution of pions.  Some have argued that this is due to 
dissipative effects.  One must be careful to not simply assume lack of full equilibration wherever
the models do not agree with data.
Also, the lifetime for the system implied by the two particle correlation (Hanbury-Brown Twiss) are 
essentially the same as at lower energies.  Perhaps this is a complex influence of late 
stage resonance gas and re-scattering, but this HBT puzzle needs to be resolved before one might
draw more detailed conclusions about the equation of state.


Agreement with hydrodynamic calculations is a strong indicator of thermalization of the system.  
However, so far no calculation is able to in detail understand the dynamics of how such fast
thermalization, less than $1-2~fm/c$, is possible.  One complication in interpreting the $v_{2}$ result 
is the balance between the equation of state and degree of thermalization.  
One can imagine that as one increases the pressure of the
system toward the Stefan-Boltzmann limit, the collective motion $v_{2}$ should increase.  However, as
the pressure is increasing, the interactions are actually decreasing in strength.  Thus, as one has
a more weakly coupled system, the time for equilibration (when one can apply the pressure) is
getting longer.  Since the system lifetime is quite short, of order $10~fm/c$, it might be that the $v_{2}$ 
would start to decrease.  In fact, if one went to a non-interacting gas limit, despite very large
pressure in equilibrium, the system would never approach equilibration and thus have no $v_{2}$.  
A non-interacting gas cannot generate elliptic flow.  Thus, even though a quark-gluon plasma will never
truly be non-interacting, there is a competition between equilibration and pressure.  Interesting
comparisons of $v_{2}$ with lower energy data may help shed light on this question.

Also, the data appear to favor an equation of state with a soft-point corresponding to a large
latent heat in a phase transition as predicted via lattice QCD.  However, any strong conclusion needs
a more systematic check on the uniqueness of the initial conditions, thermalization time, level of
equilibration and equation of state.

\section{Probes of the Medium}

Normally we would characterize an electromagnetic plasma by sending a well 
calibrated probe through the medium and observing its interactions.  Is the 
plasma transparent or opaque to our probe?  In nuclear and particle 
physics there is no feasible mechanism to aim a probe to intersect at the space 
and time where the particle collision occurs.  Therefore all our probes of the 
medium must be internally generated.

A unique feature at RHIC over lower energy experiments is that there is an appreciable 
rate of hard (high $Q^{2}$) parton-parton scattering embedded in proton-proton and 
gold-gold reactions.  These hard scatterings can be calculated within perturbative 
QCD, and are thus calibrated.  For example, in proton-proton reactions one can 
determine the yield of high transverse momentum neutral pions by assuming co-linear 
factorization, universality of parameterized structure functions and fragmentation 
functions, and pQCD determined parton-parton scattering cross sections.
\begin{equation}
{{d\sigma^{\pi^{0}}_{pp}} \over {dyd^{2}p_{T}}} = K  \sum _{abcd} \int dx_{a} dx_{b}f_{a}(x_{a},Q^{2})f_{b}(x_{b},Q^{2}){{d\sigma}\over{dt}}(ab \rightarrow cd){{D^{0}_{h/c}} \over {\pi z_{c}}}
\end{equation}
One finds that these calculations agree with the PHENIX neutral pions measured 
at 90 degrees from $p_{T} \approx 2-13$ $GeV$, at the level $\pm 50$\%.\cite{phenix_pppizero}  
This agreement is at the level of the uncertainties due to the scale chosen for the 
perturbative expansion and variations in gluon fragmentation function 
parameterizations.  

It should be noted that pQCD is well tested 
and calculationally under control compared to non-perturbative physics.  However, there are some important
disagreements.   For example beauty production differs form NLO pQCD by a factor of 2-3 
(though slightly reduced in recent years)\cite{bphys} and direct photons 
differ from NLO pQCD at lower energies and moderate $p_{T}$.\cite{D0_photon}  
At RHIC at the lowest $p_{T}$ for the neutral pions ($p_{T} \approx 2-4~GeV$) one may really not expect 
pQCD to be accurate since the coupling 
$\alpha_{s}$ at this $Q^{2}$ is not so small and the NLO calculation may miss important 
contributions to the scattering.  

\subsection{Experimental Observations}

One can extend these calculations to nuclear reactions by 
including the scaling by the nuclear thickness function.  If the parton scatterings 
are point-like, then they should simply scale with the nuclear thickness.  
Modifications from this scaling may indicate nuclear modifications in the initial 
state parton distribution functions or modifications in the final state fragmentation 
(medium effects).  

\begin{figure}[ht]
\centerline{\epsfxsize=4.1in\epsfbox{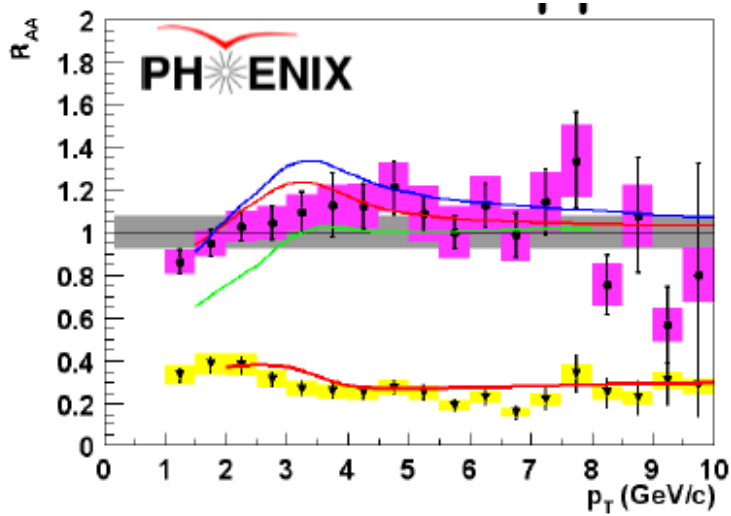}}
\caption{PHENIX data on $R_{AA}$, the ratio of neutral pion yields in $AA$ reactions
relative to $pp$ reactions scaled by the nuclear thickness, shown as the lower triangles.  
The upper data points are $R_{dA}$ comparing yields in deuteron-gold reactions to proton-proton.\label{fig_raa}}
\end{figure}

What about in heavy ion reactions?  PHENIX measurements of neutral pions in 
peripheral (large impact parameter events) agree very well from 
$p_{T} \approx 2-8$ $GeV$ with the proton-proton result including the calculated 
nuclear thickness scaling.  However, for central gold-gold reactions (small 
impact parameter) the heavy ion data is suppressed by a factor of 4-5 relative to expectations
as shown in Figure~\ref{fig_raa}.\cite{phenix_raa}  
There are four possible explanations for this large scaling violation.  

(1) One is that the initial
density of partons in the incoming nuclear wavefunction is suppressed.  This so called initial-state effect
could be the result of nuclear shadowing or saturation where the number of partons in the PDF is
reduced.  The expected level of nuclear shadowing does not explain such a large suppression, but there
are more extreme shadowing proposals where the gluon density saturates as described in color glass
condensate models.  

One prediction of a large initial-state suppression is that some suppression should remain if one 
studies these high $p_{T}$ processes in proton-gold or deuteron-gold reactions.  The gold nucleus
would still have a suppressed low x partonic content.  However, control experiment deuteron-gold
data from all four RHIC experiments shows no such suppression for hadrons at 90 degrees to the reaction,
as shown with PHENIX data in Figure~\ref{fig_raa}.

\begin{figure}[ht]
\centerline{\epsfxsize=4.1in\epsfbox{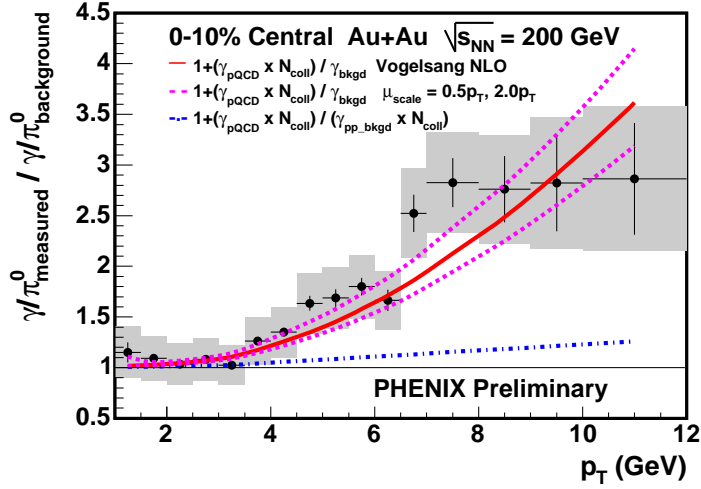}}
\caption{PHENIX preliminary direct photon results in central gold-gold reactions. 
The data is plotted as a ratio of photons to neutral pions for data compared to
simulation (including no direct photons).   The observed excess is consistent
with NLO pQCD expectations for direct photons scaled by the nuclear thickness, but
with neutral pions suppressed as previously measured by PHENIX.\label{fig_photon}}
\end{figure}

Another key check of initial-state effects is the observation of direct photons (gluon-Compton produced).  
In the direct photon case, the co-linear factorization 
equation still applies except the photon does not interact in the medium and has no
fragmentation.  Thus it is mostly sensitive to the initial PDF's and the gluon-quark 
scattering cross section.  PHENIX has measured all photons and then subtracted off photons 
from $\pi^{0}$ and $\eta$ decays.  The remaining direct photon yield is quite consistent 
with NLO pQCD calculations scaled by the nuclear thickness function as shown in 
Figure~\ref{fig_photon}.\cite{justin}  Additionally, preliminary results from the PHENIX experiment
show that the total charm quark cross section appears to scale with the nuclear thickness.\cite{kelly}
Since charm is dominantly produced in initial hard reactions and then conserved through the
medium evolution, it is quite sensitive to any modification to the incoming nuclear flux
of gluons.

Note that the deuteron-gold results and direct photon and charm results do not rule out any
initial state effect (of the order 20-30\%); however, they do rule out an initial-state effect
as being the dominant contributor to the suppression seen in gold-gold reactions.

(2) Two is that the hard scattered partons interact in the medium which modifies the energy
distribution of resulting hadrons.  
Partons will multiple scatter with other color charged objects in medium and thus will lose 
energy via induced gluon radiation (bremsstrahlung).  As the leading parton and all 
the additional radiated gluons must eventually hadronize, the experimental observable 
is a softening of the fragmentation function.  Most calculations of the angular 
distribution of the final hadrons indicate that the jet distribution will be somewhat broader 
since the radiated gluons can have of a wider angular distribution.  

(3) Three is that the pQCD calculated hard scattering cross section is incorrect.  
This possibility seems quite unlikely given the agreement of the NLO pQCD calculation 
with the proton-proton data.  In addition, above $p_{T}>10~GeV$ there should be no remaining 
soft contribution to the neutral pions.

(4) Four is that there is a complete breakdown of co-linear factorization and universality.
This is a possibility not often discussed that one should keep in mind.  In this dense color 
environment there may be much more complex couplings that violate the factorization altogether.

Here we will focus more on the second scenario of medium induced modification.
Calculations using perturbative QCD give quite good agreement with 
experimental data and imply a gluon density of order $dN_{g}/dy = 1100$ for central 
gold-gold reactions.\cite{glv,wang}  
These calculations require some additional input on 
initial state multiple scattering, shadowing and re-absorption of gluons to 
completely describe the data.  It is critical to note that the energy density implied by 
these calculations is of the same order as the energy density needed to drive the
hydrodynamic calculations into agreement with the experimental $v_{2}$ data.

An excellent test of this theory is in doing the same calculation for heavy quarks.  
At intermediate $p_{T}$ the light quarks are moved at almost the speed of light, but 
heavy quarks have a much lower velocity.  Thus there is an expected suppression of radiation 
in a forward cone called the ``dead-cone'' effect.  This means that heavy quarks should 
not be as effected by induced gluon radiation.  Data from the PHENIX experiment appears 
to support this ``lack of suppression'' but more statistics at higher $p_{T}$ are needed.\cite{kelly}  
An additional idea is to look for three jet events in the LHC heavy ion program with ATLAS and CMS.\cite{nagle_3jet}  
Since at least one jet must be a gluon, it will have a larger color coupling
to the medium and lose substantially more energy.

It will be extremely useful to have some theoretical systematic errors applied to these calculations.
Originally it was thought that in the presently available RHIC range $p_{T}<15$ $GeV$, a reliable quantitative 
prediction of quenching cannot be made due to the soft singularity that causes 
instability of the pQCD description.  In one formalism for these calculations\cite{glv}, they assume that 
no gluon modes propagate below the plasma frequency.  This provides a natural 
scale for the infrared cutoff. 
However, this would also then be true for the 0th order radiation and the
normal fragmentation process would be also be modified.  
There have been attempts to include this for heavy charm and beauty quarks, 
but not for light quarks.\cite{magdelena}

\begin{figure}[ht]
\centerline{\epsfxsize=3.1in\epsfbox{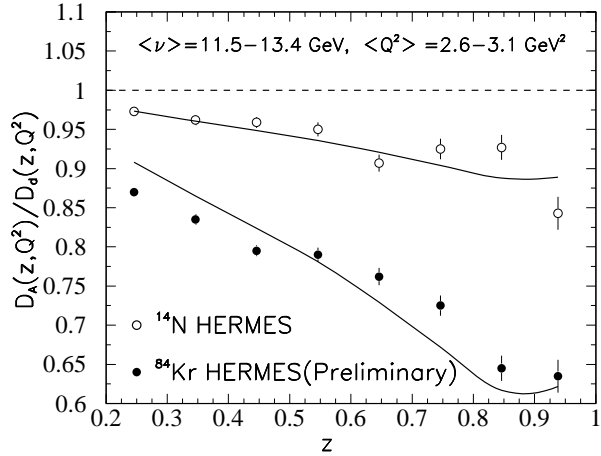}}
\caption{HERMES data comparing the relative fragmentation probability D(z) in DIS reactions
on different thickness nuclear targets..\label{fig_hermes}}
\end{figure}

The above calculations assume that the interaction with the medium is purely
partonic.  Certainly at high $p_{T}$ simple formation time considerations make it
unlikely that hadrons are formed in medium.
One interesting piece of experimental data comes from deep inelastic 
scattering on nuclei by the HERMES experiment.  They measure electrons (positrons) 
exchanging a photon with a quark in a nucleus.  By the deflection of the lepton 
one knows the energy given to the struck quark.  The quark then propagates through 
the remainder of the nucleus and can have interactions in this ``cold'' or normal 
nuclear medium.  They observe a large suppression of high z fragmentation hadrons.
They observe that larger nuclei show larger suppression as shown in Figure~\ref{fig_hermes}.

The data have been interpreted in the
same manner as heavy ion reactions, i.e. parton multiple scattering and induced radiation 
softening the fragmentation function.\cite{wang_hermes}  
It should be noted that if the parton is of sufficient 
energy (short enough wavelength), then it probes the individual partons inside of the 
nucleon constituents of the nucleus.  Thus this energy loss is not sensitive to whether 
the partons are deconfined as one might expect in a heavy ion created media or confined, 
as one knows is the case for a normal nucleus.  

There is an alternative interpretation of the HERMES data in which the hadrons are formed in medium and
high z fragments are suppressed by hadronic interactions with the nucleus.  
The observation of a large hadron species dependence of the suppression is a challenge to the
partonic energy loss picture.  One important question is whether the time scale for the parton to
come on-shell and hadronize is the same in these reactions as at RHIC.

\subsection{Jet Observations}

Jet correlations are an important test in determining 
what type of final state medium effect may be at play.  At RHIC, the detectors are 
not designed for complete jet energy reconstruction with large coverage 
electromagnetic and hadronic calorimetery (as in CDF, D0, ATLAS, CMS).  In part this 
is from budget considerations, in addition to the fact that the soft particle 
backgrounds at RHIC dominate jets within a typical jet cone size for jets less 
than $40~GeV$.  

However, the experiments can characterize jets and verify that 
high $p_{T}$ hadrons are  the result of parton-parton scattering.  STAR and 
PHENIX measure the azimuthal angular distribution of all particles relative 
to a high $p_{T}$ trigger particle.  One observes a clear ``near-angle'' peak 
from other fragmentation products of the jet and an ``away-angle'' peak from 
fragmentation of the partner scattered parton.  The away side peak is broader 
due to acceptance limitations in the longitudinal direction (rapidity) and energy 
imbalance of the two jets from final state radiation.  

\begin{figure}[ht]
\centerline{\epsfxsize=3.1in\epsfbox{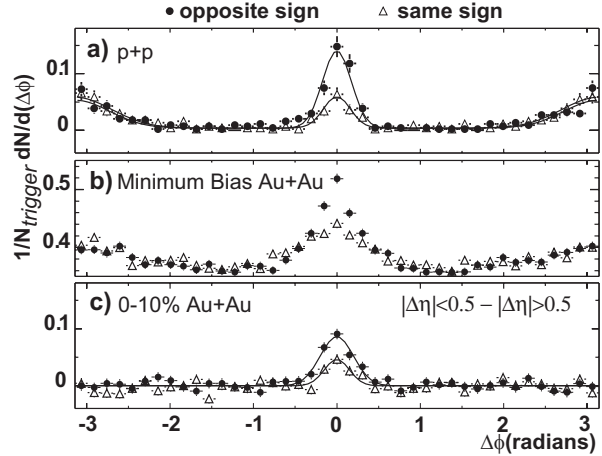}}
\caption{Azimuthal distributions of same-sign and opposite sign pairs for a) p+p
b) minimum bias Au+Au and c) background subtracted central Au+Au collisions.  All
correlation functions require a trigger particle with $4<p_{T}^{trig}<6~GeV/c$ and 
associated particles with $2<p_{T}<p_{T}^{trig}~GeV$. 
\label{fig_starjet}}
\end{figure}

In proton-proton reactions, this back-to-back angular correlations is shown in 
Figure~\ref{fig_starjet}.\cite{starjet1}  
In central gold-gold reactions, the away-side jet correlated hadrons disappear.  
As seen in Figure~\ref{fig_starjet}, the near side distribution is of similar strength 
to that in proton-proton, but the away side correlation is completely gone.  Note that 
this analysis has trigger particles from $4-6~GeV$ and only includes associated particles 
with $p_{T}>2~GeV$.  Conservation of energy and momentum says that the away side jet 
cannot really disappear.  Where has the energy gone?  One schematic way to think about 
the back-to-back jets is that if the medium somehow suppresses jet products, 
then a high $p_{T}$ trigger particle biases the ``near-angle'' jet to have come from 
the surface of the medium (thus less medium suppression).  If the ``near-angle'' is 
biased toward a short path through the medium, then the ``away-angle'' is biased 
toward a long path through the medium.  
Perhaps partons lose energy in medium, and thus the fragmentation products are all 
below $p_{T}=2~GeV$.  Perhaps some type of multiple scattering causes a broadening 
of the angular distribution that appears to remove the correlations.  My particular 
favorite idea is that we might form black holes at RHIC, which though evaporating 
quickly via Hawking radiation, might absorb the jet.  Though this is not a sound 
physics idea, it definitely gets lab management's attention.  

\begin{figure}[ht]
\centerline{\epsfxsize=3.1in\epsfbox{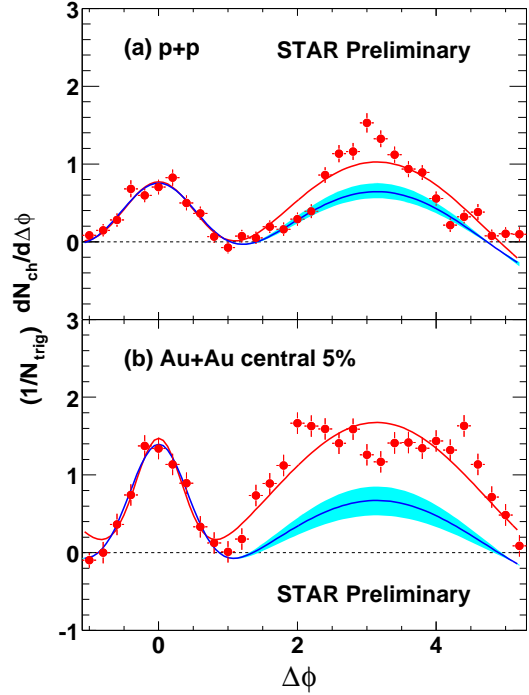}}
\caption{Preliminary results from the STAR experiment on azimuthal distributions of
hadrons in proton-proton and gold-gold reactions.  The threshold for inclusion
of hadrons in the correlation distribution is $p_{T}>200~MeV$.
\label{fig_starjet2}}
\end{figure}

The answer to this question of the missing energy was partially given at the Quark Matter 2004 conference.  
PHENIX showed preliminary results with a lower $p_{T}$ threshold
that reveal a substantial broadening of the ``away-angle'' 
jet as one compares proton-proton, peripheral gold-gold and central gold-gold reactions.\cite{phenixjet}
The broadening is more than a factor of two.  The STAR experiment showed 
preliminary results that with greater statistics and reducing the $p_{T}$ cut on the 
correlated hadrons all the way down to $200~MeV$, they then see the ``away-angle'' 
correlation.\cite{starjet2}  
Though the distribution does not appear jet-like, it answers the question 
of what happened to the energy of momentum.   One exciting observation is that the 
broadened, energy reduced jet fragmentation products end up with a momentum of order 
$500~MeV$.  This is not so different from the $<p_{T}>$ of the bulk medium.  Thus, one 
might speculate that the jet energy has been completely thermalized in medium if it 
has a long enough path through the medium.

\section{Hadronization}

\begin{figure}[ht]
\centerline{\epsfxsize=3.1in\epsfbox{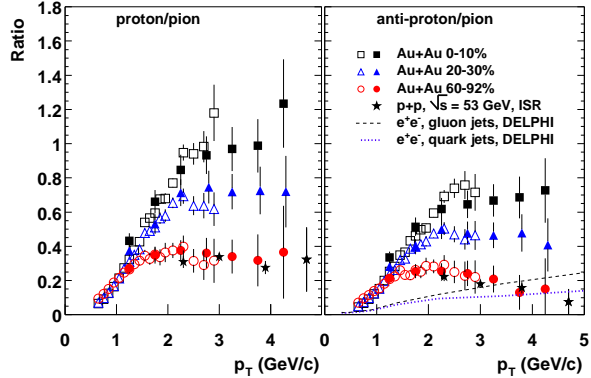}}
\caption{PHENIX ratio of (anti) protons to pions as a function of $p_{T}$ for
various centralities.\label{fig_ratios}}
\end{figure}

Another unique observation at RHIC is the particle ratios at intermediate $p_{T} \approx 2-5~GeV$.  
The jet correlation measures indicate that this momentum range has large contributions from
jet fragmentation; however PHENIX and STAR observe a large enhancement of the (anti) baryon to meson ratio
in this transverse momentum range, as shown in Figure~\ref{fig_ratios}.\cite{phenixratio}
This antibaryon and baryons excess is very different from the jet fragmentation expectation where
the ratio of baryons to mesons is always less than one.  The parton energy loss picture predicts that the
hadrons are still produced from jet fragmentation.  A recent proposal is that these intermediate baryons
and mesons have a large contribution from valence quark recombination or coalescence rather than
fragmentation.\cite{recomb}  
Since to form a baryon one needs three co-moving quarks instead of two for a meson, there
is an additional $p_{T}$ push.  These models must be confronted with more data on jet correlations for
the mesons and baryons and detailed multi-strange baryons data.  One must also distinguish checks on 
recombination as a hadron forming mechanism and the underlying distribution of partons that contribute to 
this recombination.

%

\section{Conclusions}



The relativistic heavy ion program has passed the first key test.  The nuclear physics community is capable of constructing
and running world class ``high-energy'' type experiments and re-constructing the physics
from the 10,000 particle debris.  The second phase has arrived.  Observations have been made of a very
dense gluonic medium with strong pressure built up.  The gluon density is above the predicted
phase transition level and behaving with characteristics of a fluid.  The third phase is next:
heavy quarkonia measurements to test color deconfinement, low mass vector mesons for parton correlations
in the plasma, and future comparisons with the Large Hadron Collider heavy ion program.

We have determined experimentally that a volume of matter is created the
size of a gold nucleus that is equilibrated with an initial energy density of order
ten times nuclear matter density.  If you combine this with the lattice results, you reach
the obvious conclusion that we have created a quark-gluon plasma.  However, we do not yet
have, in my opinion, compelling experimental evidence of some of the expected unique plasma properties 
predicted by QCD.   Thus, we know that we have created the quark-gluon plasma, 
but at present do not have enough information to declare a discovery - which necessitates conclusive evidence of some
unique plasma features.  This is what the future holds.

\section{Acknowledgments}

I thank the Winter Institute organizers for inviting me to give these
lectures and for providing an excellent atmosphere for exchange of ideas.
I also acknowledge useful discussions with Peter Steinberg and Tom DeGrand.

This proceedings is dedicated to my daughter Madeleine Rose Nagle, 
and the memory of my father John David Nagle.

%
%
%
%

\end{document}